# Advancing Spatio-temporal Storm Surge Prediction with Hierarchical Deep Neural Networks

**Saeed Saviz Naeini[1], Reda Snaiki[1,*], Teng Wu[2]**

[1] *Department of Construction Engineering, École de Technologie Supérieure, Université du Québec, Montréal, Canada*

[2] *Department of Civil, Structural and Environmental Engineering, University at Buffalo, Buffalo, NY 14260, USA*

*\*Corresponding author. Email:* reda.snaiki@etsmtl.ca

**Abstract:** Coastal regions in North America face significant threats from storm surges caused by hurricanes and nor'easters. While traditional numerical models offer high-fidelity simulations, their computational costs limit their practical use for real-time predictions and risk assessments. Accordingly, the deep learning techniques have been recently developed for efficient simulation of time-dependent storm surge with inputs of storm parameters. To resolve the small scales of storm surge in both time and space over a long duration and a large area, these simulations typically need to employ oversized neural networks that struggle with the accumulation of prediction errors over successive time steps. To address these challenges, this study introduces the hierarchical deep neural network (HDNN) technique combined with a convolutional autoencoder to accurately and efficiently predict storm surge time series. The convolutional autoencoder is used to reduce the dimensionality of the storm surge data, making the learning process more concise and efficient. The HDNNs then map storm parameters to the low-dimensional representation of storm surge, allowing for sequential predictions across different time scales. Specifically, the current-level neural network is utilized to predict future states with a relatively large time step, which are passed as inputs to the next-level neural network for smaller time-step predictions. This process continues sequentially for all time steps. The simulation results from different-level neural networks across various time steps are then stacked to acquire the entire time series of storm surge. The simulated low-dimensional representations are finally decoded back into storm surge time series. The proposed model was trained and evaluated using synthetic data from the North Atlantic Comprehensive Coastal Study (NACCS). Results demonstrate its excellent performance to effectively handle high-dimensional surge data while mitigating the accumulation of prediction errors over time, making it a promising tool for advancing spatio-temporal storm surge prediction.

## 1. Introduction

Storm surge is a coastal hazard fueled by powerful winds and low atmospheric pressure associated with tropical and extratropical cyclones (Bij de Vaate et al., 2024). Due to climate change, increasing storm frequency and intensity, along with rising sea levels, will exacerbate the effects of storm surge, leading to severe property damage, loss of life, and disruption of essential services (Calafat et al., 2022; Gori et al., 2022; Snaiki and Wu, 2020). To mitigate these impacts, coastal communities must implement effective mitigation measures such as coastal protection, land use planning, early warning systems, and emergency preparedness. Accurate storm surge modeling is crucial for achieving this goal. Existing storm surge prediction models usually fall into four main categories, namely statistical, empirical, numerical, and data-driven. While efficient, empirical and statistical models are often limited to specific areas and may not accurately capture nonlinear dynamics (Jelesnianski, 1972; Sztobryn, 2003; Thomas and Dwarakish, 2015). High-fidelity numerical models, like ADCIRC, offer superior accuracy but are computationally expensive, making them unsuitable for applications like probabilistic hazard assessments and near-real-time predictions. Data-driven techniques, which learn from numerical, experimental, or field measurement data, offer a promising alternative for rapid storm surge modeling (Saviz Naeini and Snaiki, 2024a).

Data-driven techniques have emerged as powerful tools for predicting storm surge, offering near-instantaneous forecasts based on the relationship between storm parameters and resulting coastal water levels. These techniques are trained to identify underlying patterns and dependencies, enabling accurate predictions of storm surge. Numerous studies have explored various data-driven approaches for peak storm surge prediction. Specifically, Jiang et al. (2024) categorized these techniques into kriging (e.g., Jia and Taflanidis, 2013; Jung et al., 2023; Kijewski-Correa et al., 2020; Kyprioti et al., 2021; Plumlee et al., 2021; Zhang et al., 2018), support vector machines (e.g., Al Kajbaf and Bensi, 2020; Hashemi et al., 2016; Rajasekaran et al., 2008), and artificial neural networks (e.g., Ayyad et al., 2022; Bajo and Umgiesser, 2010; Chen et al., 2012; Lee et al., 2021; Lockwood et al., 2022; Pachev et al., 2023; Ramos-Valle et al., 2021; Saviz Naeini and Snaiki, 2024b). A significant challenge in storm surge prediction lies in the need to cover extended coastal regions. Traditional approaches, often limited to specific locations, struggle to capture the complex spatial and temporal dynamics of storm surges. To address this limitation, dimensionality reduction techniques have been introduced. By reducing the dimensionality of the data while preserving essential information, these techniques enable the application of reduced-order models to simulate storm surge over larger areas (e.g., Jia and Taflanidis, 2013; Kyprioti et al., 2021; Saviz Naeini and Snaiki, 2023). Nonlinear dimensionality reduction methods, such as autoencoders, are particularly well-suited for capturing complex nonlinear relationships within the data. While most existing methods separate the dimensionality reduction process from the surrogate model used to simulate storm surge, Saviz and Snaiki (2024b) introduced a novel approach to simultaneously identify the latent space through dimensionality reduction and train a data-driven model within this space, significantly enhancing storm surge prediction accuracy.

While most studies have focused on predicting the peak storm surge, relatively few have addressed the simulation of time-series storm surge (e.g.,Giaremis et al., 2024; Igarashi and Tajima, 2021; Kim et al., 2015; Qin et al., 2024; Xie et al., 2023). Traditional data-driven approaches struggle to capture the inherent complexities of storm surge, which is a highly dynamic phenomenon. The multiple spatial scales introduce additional complexity to the modeling process, making it challenging to simultaneously capture both spatial and temporal aspects of storm surge. Kyprioti et al. (2023) and Adeli et al. (2023) are examples of studies that have attempted to address both spatial and temporal aspects of storm surge prediction. Kyprioti et al. (2023) used Gaussian Process techniques for large-scale time-series simulation, while Adeli et al. (2023) combined CNN and LSTM models for spatio-temporal modeling of storm surge in the Coastal Texas region. Despite advancements in machine learning, accurate simulations of time-dependent problems remain challenging due to the accumulation of prediction errors, exploding/vanishing gradients, computational complexity, and overfitting.

In this study, a hybrid machine learning model is proposed to accurately and efficiently predict storm surge time series. This model employs the hierarchical deep neural networks (HDNNs) combined with a convolutional autoencoder, leveraging the strengths of both techniques. Specifically, the convolutional autoencoder reduces the dimensionality of the storm surge data, making the learning process more concise and efficient. The HDNNs then map storm parameters (central pressure deficit, radius of maximum winds, storm latitude, and storm longitude) to this low-dimensional representation of storm surge, allowing for sequential predictions across different time scales. The simulated low-dimensional representations are finally decoded back into storm surge time series. The proposed model was trained and evaluated using synthetic data from the North Atlantic Comprehensive Coastal Study (NACCS), covering large coastal areas of New York and New Jersey. Results demonstrate its excellent performance to effectively handle high-dimensional surge data while mitigating the accumulation of prediction errors over time, making it a promising tool for advancing spatio-temporal storm surge prediction.

## 2. Methods

### 2.1. Problem statement

The governing equations of storm surge represented as partial differential equations (PDEs), are typically solved using advanced numerical techniques, where time-stepping algorithms are designed to accurately resolve the hydrodynamics. To achieve the high level of accuracy, it is essential to select a very small time-step size. However, this requirement significantly increases the computational cost. As a result, applying these high-fidelity models in real-time predictions or risk assessments can be quite challenging.

To illustrate the complexities involved in simulating the storm surge process in an $n$-dimensional space, consider the following dynamical system that predicts the storm surge height $\eta(t) \in \mathbb{R}^n$ using a set of hydrodynamic equations $f$:

$$\dot{\eta}(t) = f(\eta(t), \theta(t), t) \tag{1}$$

where $t$ is time and $\theta(t)$ includes storm parameters. Four storm parameters considered in this study are expressed as follows (Adeli et al., 2023):

$$\theta(t) = [C_p(t); R_{max}(t); LAT(t); LON(t)]^T \tag{2}$$

where $C_p(t)$ = central pressure deficit (hPa); $R_{max}(t)$ = radius of maximum winds (km); $LAT(t)$ = latitude of the storm eye (°); and $LON(t)$ = longitude of the storm eye (°). A time-stepping algorithm can be employed to integrate the storm surge dynamics, resulting in the storm surge response at the next time step (i.e., $t + \Delta t$):

$$\eta(t + \Delta t) = \boldsymbol{F}(\eta(t), \theta(t), \Delta t) \tag{3}$$

where $\boldsymbol{F}$ integrates the function $f$ between $t$ and $t + \Delta t$, typically approximated using a time-stepping algorithm such as the Runge-Kutta scheme. In this study, the hierarchical deep learning of homogeneous differential equation time-steppers, originally developed by (Liu et al., 2022), has been introduced to generate an approximation $\widehat{\boldsymbol{F}}$ of the discrete function $\boldsymbol{F}$ and generalized to account for the storm parameters $\theta(t)$.

## 2.2. Hierarchical neural networks

To address the high computational costs associated with traditional time-stepping algorithms, artificial neural networks (ANNs) or deep neural networks (DNNs) can be effectively trained to identify a function $\widehat{\boldsymbol{F}}$ which approximates $\boldsymbol{F}$ (Eq. 3) and predict time-series storm surge in real-time. However, developing a single ANN (or DNN) model to predict $\eta(t + \Delta t)$ based on the current storm surge value $\eta(t)$ can be challenging due to the accumulation of errors that occur during the training and prediction process. To overcome this issue, advanced architectures suitable for time-dependent problems, such as recurrent neural networks (RNNs) and the more sophisticated long short-term memory (LSTM) algorithms, have been proposed in the literature. Despite their advanced capabilities, these models can still encounter issues such as error accumulation, the exploding/vanishing gradient problem, computational complexity, and overfitting. In this study, a hierarchy of DNN models is instead employed. Specifically, a set of $k$ neural networks $[\widehat{F}_1, \widehat{F}_2, \dots, \widehat{F}_i, \dots, \widehat{F}_k]$ will be trained using different time steps $\Delta t_i$ such that:

$$\eta(t + \Delta t_i) = \widehat{F}_i(\eta(t), \theta(t), \Delta t_i) \tag{4}$$

The upper and lower bound indices are determined iteratively based on the specific problem to retain the most effective time-stepping neural networks. With the trained $\widehat{F}_i$ neural networks, storm surge predictions can be made through a straightforward vectorization process (Liu et al., 2022). Specifically, the neural network corresponding to the largest time step is utilized to predict future states, which are then fed as inputs into the next-level neural network. This sequential process continues for all time steps. The simulation results from the various neural networks across different time steps are subsequently combined (arranged by chronological order) to create a comprehensive time series of storm surge. This procedure is schematically illustrated in Fig. 1, showcasing three neural networks trained with three different time steps ($\Delta t_1, \Delta t_2, \Delta t_3$).

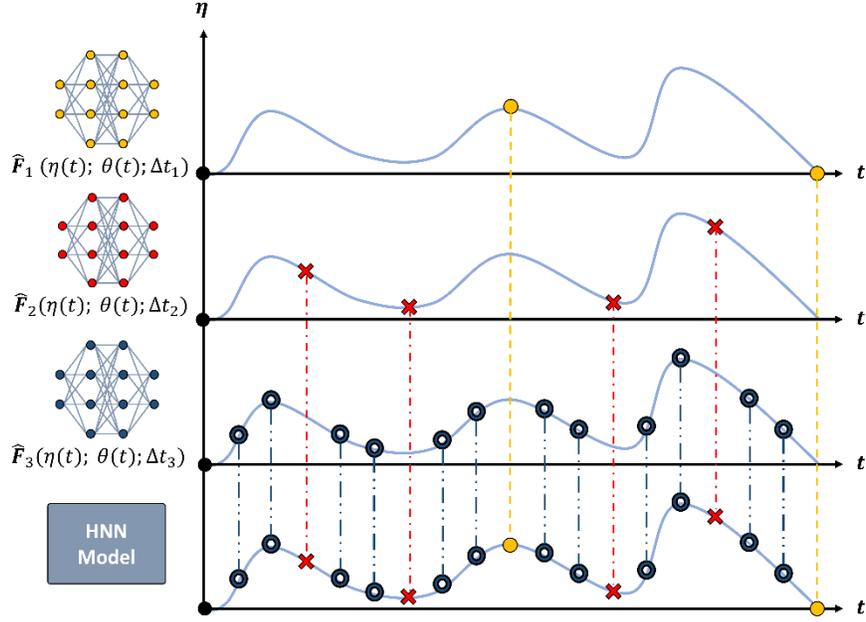

**Fig. 1.** Schematic illustration of the predication and vectorization processes of the hierarchical neural networks

As illustrated in Fig. 1, the model with the largest time step ($\hat{F}_1$) is initially employed to predict future storm surge values using the time step $\Delta t_1$. The predicted storm surge values are then fed as inputs into the next neural network, $\hat{F}_2$, which corresponds to the second largest time step ($\Delta t_2$). This network predicts future storm surge values within each interval defined by $\Delta t_1$. The process continues sequentially, predicting storm surge values within each interval defined by the predicted values at $\Delta t_2$, ultimately generating storm surge predictions with a time step of $\Delta t_3$. It is important to highlight that a key difference between the original HDNN (Liu et al., 2022) and the modified version presented here lies in the inclusion of storm parameters $\theta(t)$ during the prediction phase of each $\hat{F}_i$ model. This modification significantly enhances the model versatility, enabling its application to a wide range of storm scenarios rather than being limited to a single, specific case. Specifically, the neural network $\hat{F}_i$ leverages both the current storm surge values and pertinent storm parameters at each time step as inputs to predict the subsequent storm surge value at time $t + \Delta t_i$.

## 2.3. Dimensionality reduction

The storm surge response is typically simulated across a large spatial region, making it extremely challenging to train a single set of hierarchical neural networks for storm surge prediction of the entire area. Also, it is impractical to develop one set of hierarchical neural networks for each geographical point. To address the challenges posed by the high dimensionality of the output vector (i.e., storm surge), a dimensionality reduction technique is necessary. Numerous dimensionality reduction techniques have been proposed in the literature to identify a low-dimensional feature space from a high-dimensional dataset (Abdi and Williams, 2010; Saviz Naeini and Snaiki, 2023; Van Der Maaten et al., 2009). Once a low-dimensional space is established, reduced-order models can be applied to this representation for system approximation. In this study, the autoencoder technique is utilized which is recognized as one of the most advanced methods for nonlinear dimensionality reduction. The autoencoder consists of two main components, namely an encoder ($\varphi$) and a decoder ($\psi$). The encoder maps the input data (i.e., storm surge values across the entire region, $\eta$) into a reduced-dimensional latent representation ($z$), such that $z = \varphi(\eta)$; the decoder then reconstructs the original input data from this latent representation, resulting in $\hat{\eta} = \psi(z)$. The training process focuses on minimizing the reconstruction error between the original input ($\eta$) and the reconstructed output ($\hat{\eta}$). This objective compels the autoencoder to learn a compressed representation ($z$) that effectively captures the essential features of the input data. In contrast to linear dimensionality reduction techniques like principal component analysis (PCA), autoencoders are capable of handling complex nonlinear relationships within the data, thus facilitating more effective representation learning (Liou et al., 2014; Wetzel, 2017). However, a significant challenge associated with typical fully connected autoencoders is the large number of trainable parameters in these architectures, which can often lead to issues of intractability. An alternative approach employed here is the use of convolutional autoencoders (CAEs) (Yu and Wu, 2023). Figure 2 provides a schematic representation of a typical CAE architecture. Like standard autoencoders, CAEs are composed of an encoder and a decoder. The encoder features multiple convolutional layers designed to extract relevant features from the input data. It may also include pooling layers and filters. Conversely, the decoder comprises up-sampling layers that increase the dimensionality of the feature maps, along with convolutional layers that reconstruct the original input data from the learned features. Given the capabilities of convolutional autoencoders, this study utilizes CAEs to derive a reduced spatial representation of the high-dimensional space, specifically focusing on time-series storm surge responses across an extensive coastal area.

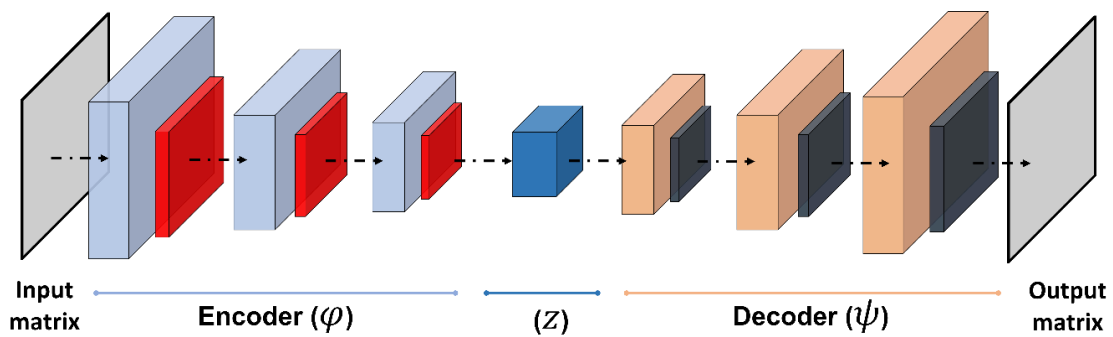

**Fig. 2.** Architecture of a typical CAE

## 2.4. Proposed scheme

This study develops a machine learning-based hybrid scheme to simulate time-series storm surge across an extensive region, as schematically illustrated in Fig. 3. By harnessing the capabilities of CAEs, a lower-dimensional latent space is first identified that captures the essential information from the original storm surge data. This reduced representation enables the implementation of HDNNs, which are trained on the latent space to map the current storm surge values, along with the storm parameters, to the corresponding storm surge values at subsequent time steps.

In the initial step, the required time series storm surge data across all points within the selected geographical region are extracted and processed for training purpose. Specifically, the data are reorganized into a 2D matrix format, where each matrix represents storm surge values across all points (referred to as save points) at a specific time step for a given storm scenario. This process results in a total of $p \times m$ matrices, where $p$ is the total number of time steps per scenario (assumed to be the same for all scenarios to simplify the implementation) and $m$ represents the number of storm scenarios. The dimensions of the 2D matrices are determined by the total number of save points within the studied geographical region. These generated matrices are then utilized to train the CAE model, as illustrated in Fig. 2.

Once the CAE model is trained, a lower-dimensional latent space is identified in the form of 1D vectors. By utilizing the CAE encoder with the preprocessed storm surge dataset as inputs (comprising a total of $p \times m$ matrices), $p \times m$ 1D vectors are generated that will be used to train the HDNNs. This generated data is reorganized into a set of input/output pairs, where the input consists of the transformed storm surge values at time step $t$ (represented as 1D vectors) along with the storm parameters (concatenated with the storm surge data). The output corresponds to the transformed storm surge values at time step $t + \Delta t_i$. In this case, a set of $k$ neural networks $[\widehat{F}_1, \widehat{F}_2, \ldots, \widehat{F}_i, \ldots, \widehat{F}_k]$ is trained with different time steps $\Delta t_i$ to predict storm surge values in the latent space. After training the HDNNs, any storm scenario can be simulated, allowing for the entire time series of storm surge in the latent space to be obtained using the vectorization process detailed in Section 2.2. The final step involves employing the decoder of the CAE model to reconstruct the high-dimensional representation of the storm surge across the entire geographical region.

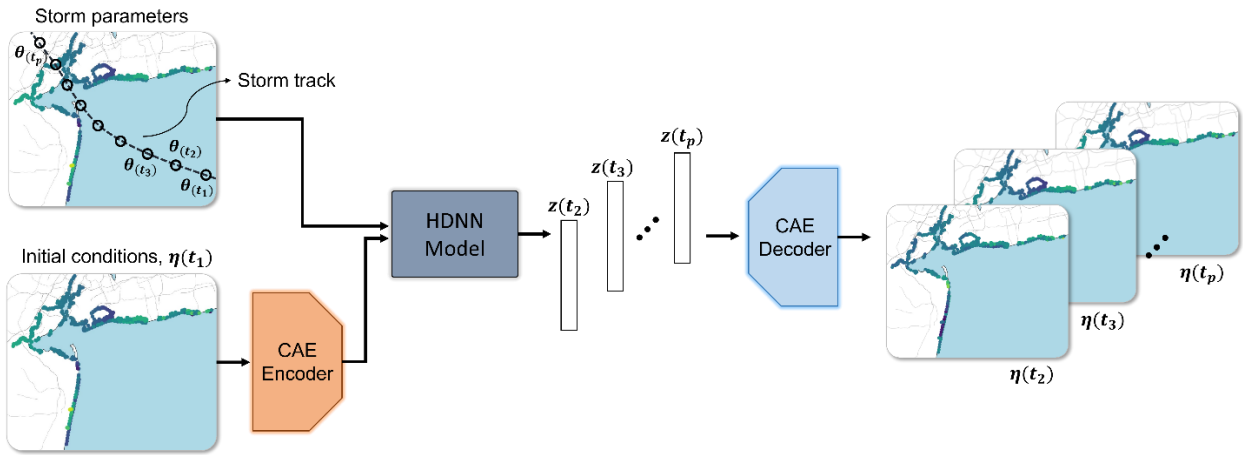

**Fig. 3.** Schematic framework of the proposed predictive model for storm surge

## 3. Case study

The case study is focused on the simulation of time-series storm surge events across various locations in New York and New Jersey, as illustrated in Fig. 4. These coastal regions are particularly vulnerable to storm surge and flooding, as evidenced by the devastating impact of Hurricane Sandy in 2013 (Coch, 2014). Given the anticipated intensification of climate change and rising sea levels, the risk of storm surge in this region is expected to increase significantly (Ayyad et al., 2023). Consequently, rapid and accurate storm surge prediction is essential. This section begins by outlining the dataset used for training and testing the proposed model. Subsequently, the training performance of both CAE and HDNN models is presented. Finally, an illustrative example demonstrates the practical application of the proposed machine learning-based hybrid scheme for simulating storm surge events.

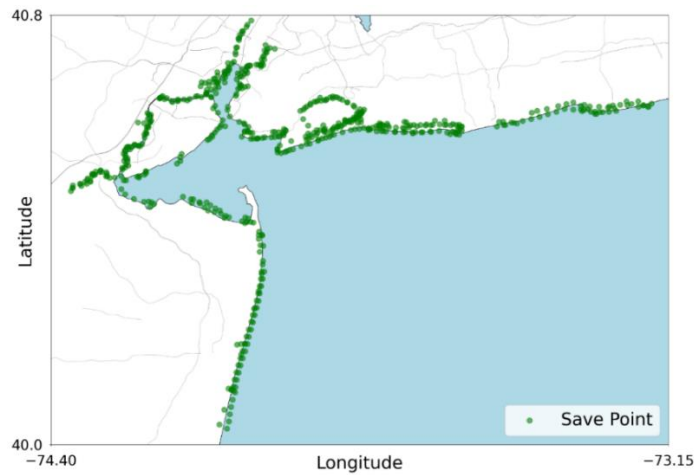

**Fig. 4.** Study area with the representation of save points

### 3.1. Database

The present study leverages the North Atlantic Comprehensive Coastal Study (NACCS) database, accessible through the Coastal Hazards System (CHS) web-tool. NACCS provides high-resolution numerical simulations of storm surge from 1050 synthetic Tropical Cyclones (TCs). These simulations are generated by coupling the ADCIRC and STWAVE models. After data preprocessing, which eliminated 19 storms with unreliable simulations, a subset of 344 TCs was selected based on their proximity to the study area (within a 200-km radius). Each of these 344 storms is represented by 125 time steps, each corresponding to a 10-minute interval. The 125 time steps are centered around the reference location defined by NACCS, which can be considered as the approximate landfall location for landfalling scenarios. The time series spans 100 steps before this point and 25 steps after it. The storm surge simulations are recorded at 578 save points within the study area.

Based on the NACCS database, two datasets are created. The first dataset, used to train the CAE model, consists of time series storm surge values organized into a matrix format as explained in Sect. 2.4. This dataset is randomly divided into training (80%), validation (10%), and testing (10%) sets to ensure robust model performance. After training the CAE model, a lower-dimensional latent space is identified as 1D vectors. The CAE encoder is then applied to the preprocessed storm surge dataset ($p \times m$ matrices), generating $p \times m$ 1D vectors. The second dataset is formed by reorganizing the generated data into input/output pairs. The input includes the transformed storm surge values at time step $t$ (1D vectors) and the storm parameters (concatenated with the storm surge data). The output corresponds to the transformed storm surge values at time step $t + \Delta t_i$. This dataset is randomly divided into training (70%), validation (15%), and testing (15%) subsets to train the HDNN models.

## 3.2. Training performance

Several metrics are used in this study to evaluate the performance of the CAE and HDNN models, namely the mean-squared error (MSE), the root-mean-squared error (RMSE) and the mean absolute error (MAE). The training, validation and testing results for both CAE and HDNN models are presented in the following sections.

### 3.2.1. Convolutional autoencoder

The primary goal of the CAE is to reduce the high dimensionality associated with the time-series storm surge simulations across the 578 save points by identifying a lower-dimensional latent space. As described in Sections 2.4 and 3.1, the data is restructured into a matrix format with dimensions 24x24 (=578). To enhance data representation and facilitate efficient model training, this data is subsequently normalized within the [0,1] range. The encoder takes a (24, 24, 1) input matrix and compresses it to a 4-dimensional encoding vector. This compression is achieved through four convolutional layers (Conv2D) with increasing filter counts (16, 32, 64, 128), each followed by a max pooling layer (MaxPooling2D). Each Conv2D layer uses a 3x3 spatial filter, and all max pooling layers have a 2x2 window, except for the third which uses a 3x3 window. The final layers are a flattening layer and a dense layer with 4 neurons. The decoder reconstructs the original image from the compressed encoding vector. It starts with a dense layer followed by a reshaping layer. Then, four convolutional transpose layers (Conv2DTranspose) with decreasing filter counts (128, 64, 32, 16) are used, each followed by an upsampling layer (UpSampling2D). Upsampling layers use 2x2 windows, except for the second, which uses a 3x3 window. ReLU activation is applied to Conv2D and Conv2DTranspose layers, except for the final decoder layer, which uses a Sigmoid function to constrain output values between 0 and 1. The model is trained using the Adam optimizer with a learning rate of 0.001, and MSE loss was used to measure the reconstruction error. Weights were initialized with a Gaussian distribution, and the model underwent 2000 epochs of training. Figure 5 provides a simplified diagram of the CAE architecture, highlighting the convolutional layers, convolutional transpose layers, flattening layers, and the 4-dimensional encoder vector.

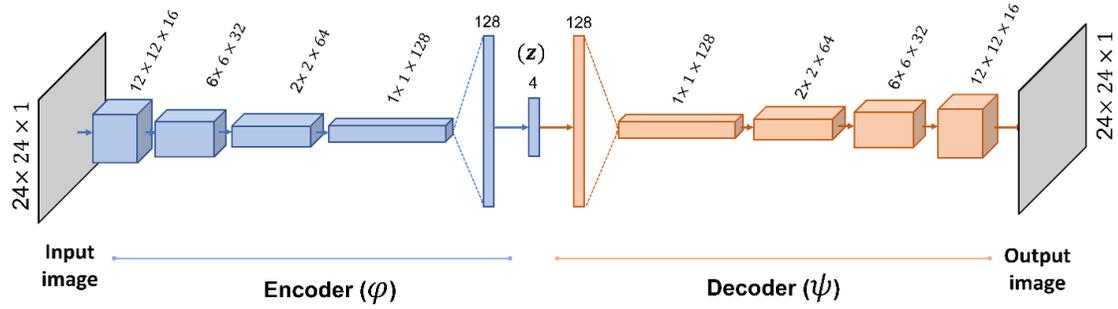

**Fig. 5.** Architecture of the proposed CAE model

The CAE model training and validation losses demonstrate its good performance as shown in Fig. 6. The training loss has converged to a low value of 0.0003, indicating effective learning of the dataset. Furthermore, the validation loss has steadily decreased to a low value of 0.0005, suggesting that the model is generalizing well to unseen data. These results collectively suggest that the CAE model has achieved satisfactory performance. In addition, Table 1 summarizes the performance of the CAE model in terms of RMSE and MAE for the training, validation, and testing sets. Based on the obtained RMSE and MAE values across all three sets, the CAE model demonstrates satisfactory performance in reconstructing the input data. Both RMSE and MAE are relatively low, indicating that the model is able to accurately approximate the original data.

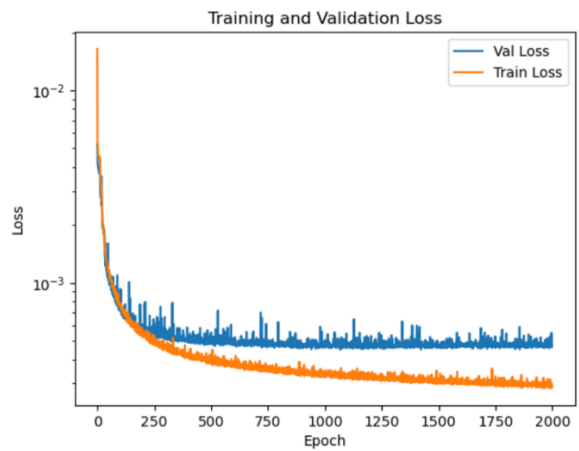

**Fig. 6.** Performance of CAE training process (loss MSE)

**Table 1.** Comparison of CAE training performance

| Metric | Training set | Validation set | Testing set |
|--------|--------------|----------------|-------------|
| RMSE   | 0.017        | 0.022          | 0.026       |
| MAE    | 0.010        | 0.013          | 0.015       |

### 3.2.2. Hierarchical deep neural networks

After identifying a lower-dimensional latent space which captures the essential information from the original storm surge data using the CAE model, the HDNNs are implemented to map the current storm surge values, along with the storm parameters (Eq. 3), to the corresponding storm surge values at subsequent time steps. The necessary data for training the HDNNs are generated and organized as described by Sect. 2.4. A set of 7 neural networks $[\widehat{F}_1, \widehat{F}_2, \widehat{F}_3, \widehat{F}_4, \widehat{F}_5, \widehat{F}_6, \widehat{F}_7]$ is trained with different time steps $\Delta t_i$ (1, 2, 4, 8, 16, 32, and 64) to predict storm surge values in the latent space. All DNN models employed the same architecture, consisting of an input layer with 8 nodes representing storm parameters and latent variables, followed by three hidden layers with 512 neurons each. The output layer had 4 neurons corresponding to the latent variables. ReLU activation was used for the hidden layers, while a linear activation function was applied to the output layer. Weights were initialized using Xavier initialization, and biases were set to zero. The models were trained using the Adam optimizer with a learning rate of 0.001, and a batch size of 128. Early stopping was implemented to prevent overfitting, terminating training if the validation loss did not improve for 50 consecutive epochs. While the maximum number of epochs was set to 2000, the actual number of epochs varied due to early stopping. Figure 7 illustrates the training and validation loss curves for each DNN model. As shown, both training and validation losses consistently decreased with increasing epochs, indicating effective learning and generalization. This observation suggests that the models were able to learn the underlying patterns in the data without overfitting.

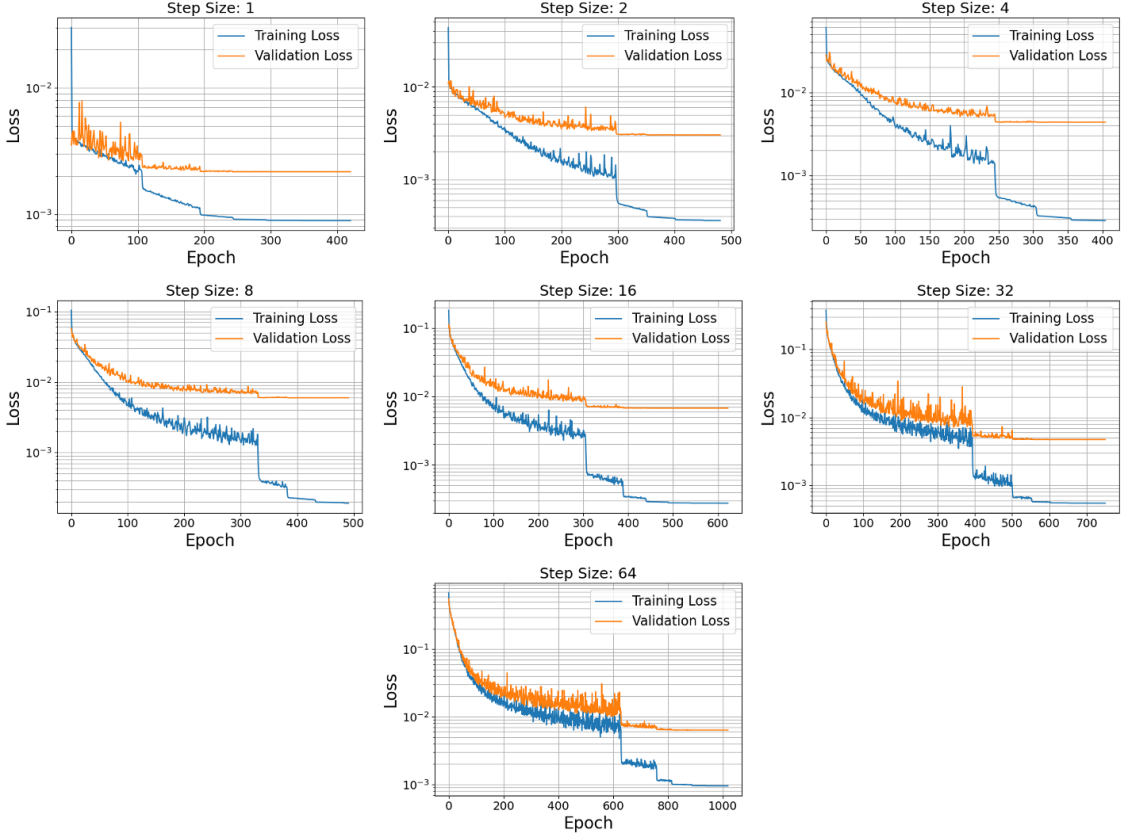

**Fig. 7.** Performance of DNN training process

The trained $\widehat{F}_i$ neural networks can be used to predict storm surges through a sequential vectorization process as proposed by Liu et al. (2022). Starting with the largest time step (Δt = 64), the corresponding neural network is used to predict future states. These predictions are then combined with the initial states and fed as input to the next-level neural network (Δt = 32). This sequential process continues through all time steps, progressively decreasing the time step (Δt = 16, Δt = 8, Δt = 4, Δt = 2, Δt = 1). The vectorization process ultimately results in a comprehensive time series of predicted storm surge values. When evaluated on the testing set, the combined DNN models achieved an RMSE of 0.050 and an MAE of 0.023. These low error values indicate good overall performance of the hierarchical model. Following the training of the HDNNs, the final step is to use the CAE decoder to reconstruct a high-dimensional representation of the storm surge across the entire geographical region. This process will be explored in detail through an application example in the next section.

### 3.3. Application

In this section, four random scenarios selected from the testing set will be investigated to assess the proposed machine learning-based hybrid scheme for predicting time-series storm surge. The tracks of the four storm scenarios are depicted in Fig. 8 with IDs 293, 402, 460, and 541. To predict the storm surge responses over the entire region (Fig. 4), the storm parameters $\theta(t)$ at each time step are needed and can be extracted from storm tracks. The initial storm surge values across the entire region needed as inputs to the HDNNs will be obtained using the values obtained from the NACCS database and then converted to the reduced latent space using the encoder of the CAE model. To illustrate the simulation of time series storm surge, two save points are selected with IDs SP03970 and SP03999 and depicted in Fig. 8.

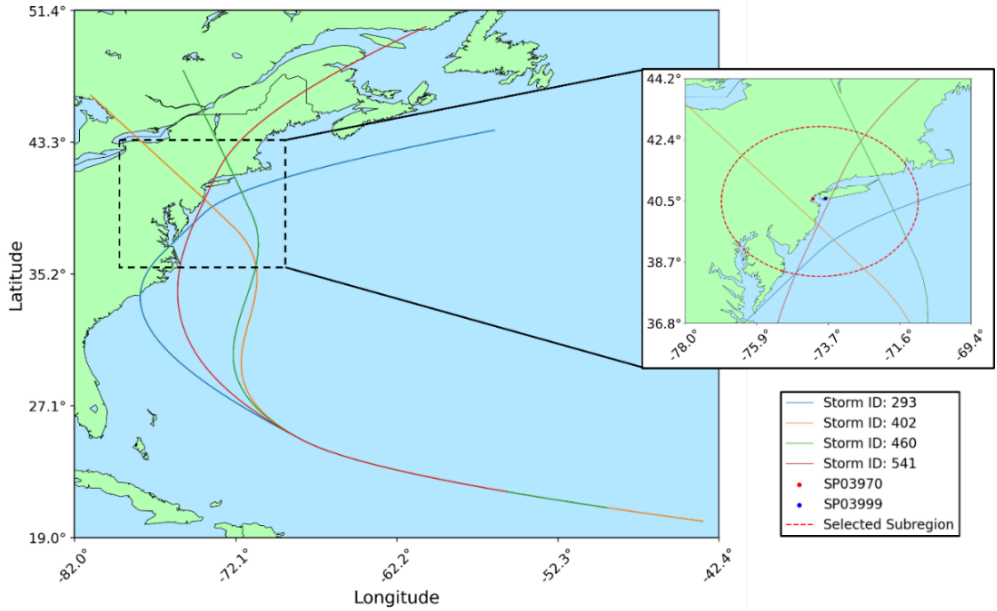

**Fig. 8.** Selected storm tracks and save points for model performance evaluation

Once the predictions have been made using the HDNN technique, the predicted values in the latent space are transformed to the high-dimensional space using the CAE decoder. The simulation results for time-series storm surge at the two selected locations with four track scenarios retrieved from the test dataset are illustrated in Fig. 9.

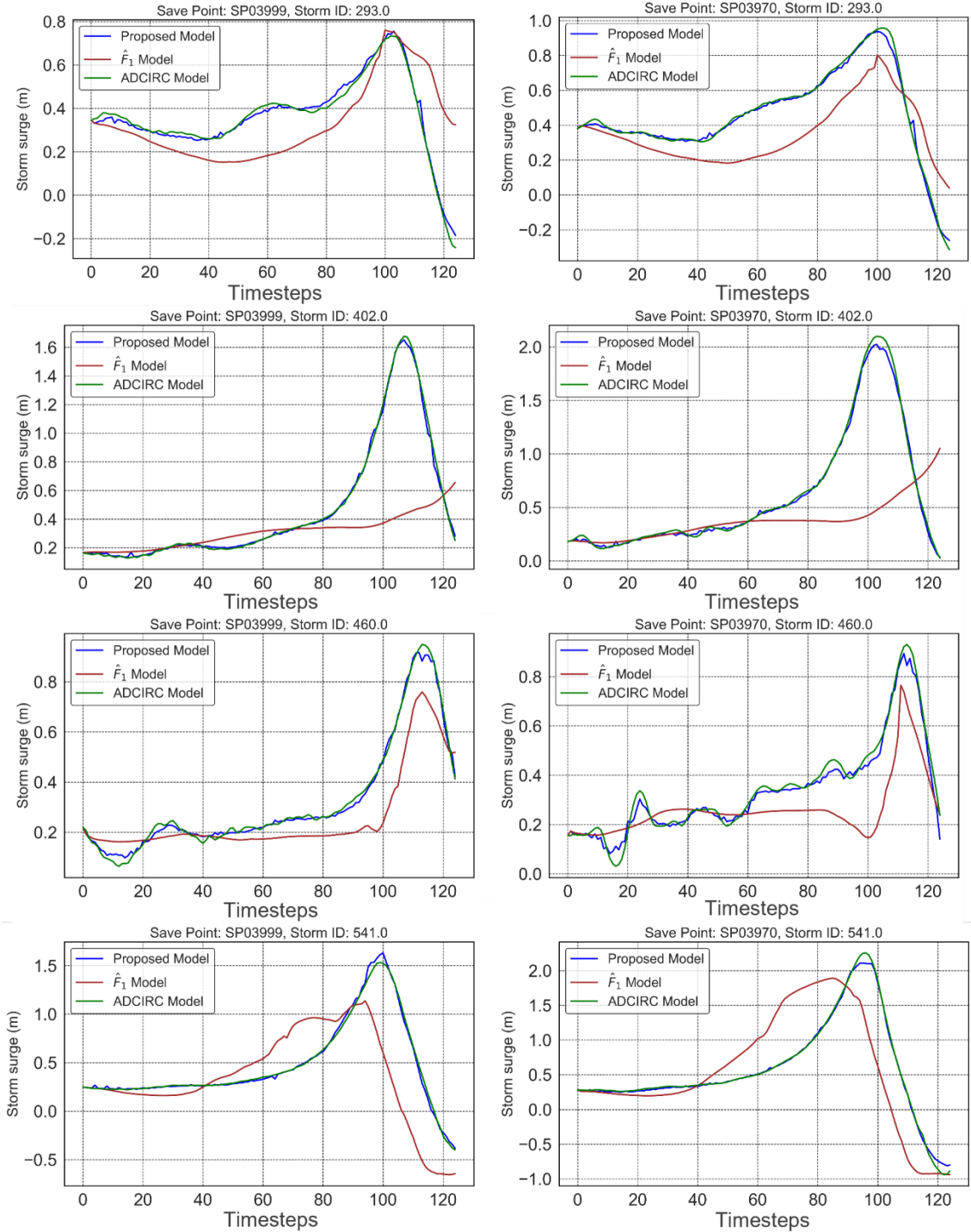

**Fig. 9.** Time-series storm surge predictions at the selected points, #SP03999 (left column) and #SP03970 (right column), given four different storm scenarios with IDs 293, 402, 460, and 541 (up to bottom) (each time step is equivalent to 10 min)

Figure 9 illustrates an excellent agreement between the simulated time-series storm surge values from HDNNs and the results from ADCIRC-based simulations for all four scenarios. This observation demonstrates the model effectiveness in accurately predicting storm surge throughout the entire time series with insignificant errors. To highlight the superiority of the proposed hybrid approach, the time-series storm surge predictions using a single DNN model with the smallest time step ($\widehat{F}_1$) were also simulated and compared. As shown in Fig. 9, the $\widehat{F}_1$ model exhibits limited accuracy, particularly in predicting far-time steps due to the cumulative accumulation of errors. This comparison clearly demonstrates the significant improvement in accuracy achieved by the HDNNs in predicting time-series storm surge.

To examine the spatial distribution of storm surge across the study region, Fig. 10 presents the storm surge distribution at time steps 60 and 100, induced by storm ID 293. By comparing the ADCIRC-based storm surge distribution with that generated by the proposed hybrid scheme, it is evident that excellent agreement is achieved at different time steps. This observation is supported by the small differences observed between the two simulations, as shown in the same figure. Specifically, the absolute value of storm surge difference is limited to less than 0.10 meters, reaching up to 9% error only in a few isolated locations at peak storm surge. The majority of locations exhibit errors significantly below this threshold. Higher storm surge values are observed at time step 100, which corresponds to the peak storm surge as depicted in Fig. 9. This increase is attributed to the storm's proximity to landfall, combined with a high central pressure deficit, which drives elevated surge levels. Additionally, the spatial distribution of storm surge reveals that certain locations are more susceptible to higher surge heights than others. This variation is primarily influenced by the proximity to the storm track, the relative position with respect to the storm, and other factors such as bathymetry, coastal shape and topography. Therefore, accurately modeling storm surge is crucial for rapidly identifying these critical regions, enabling effective coastal management and disaster preparedness.

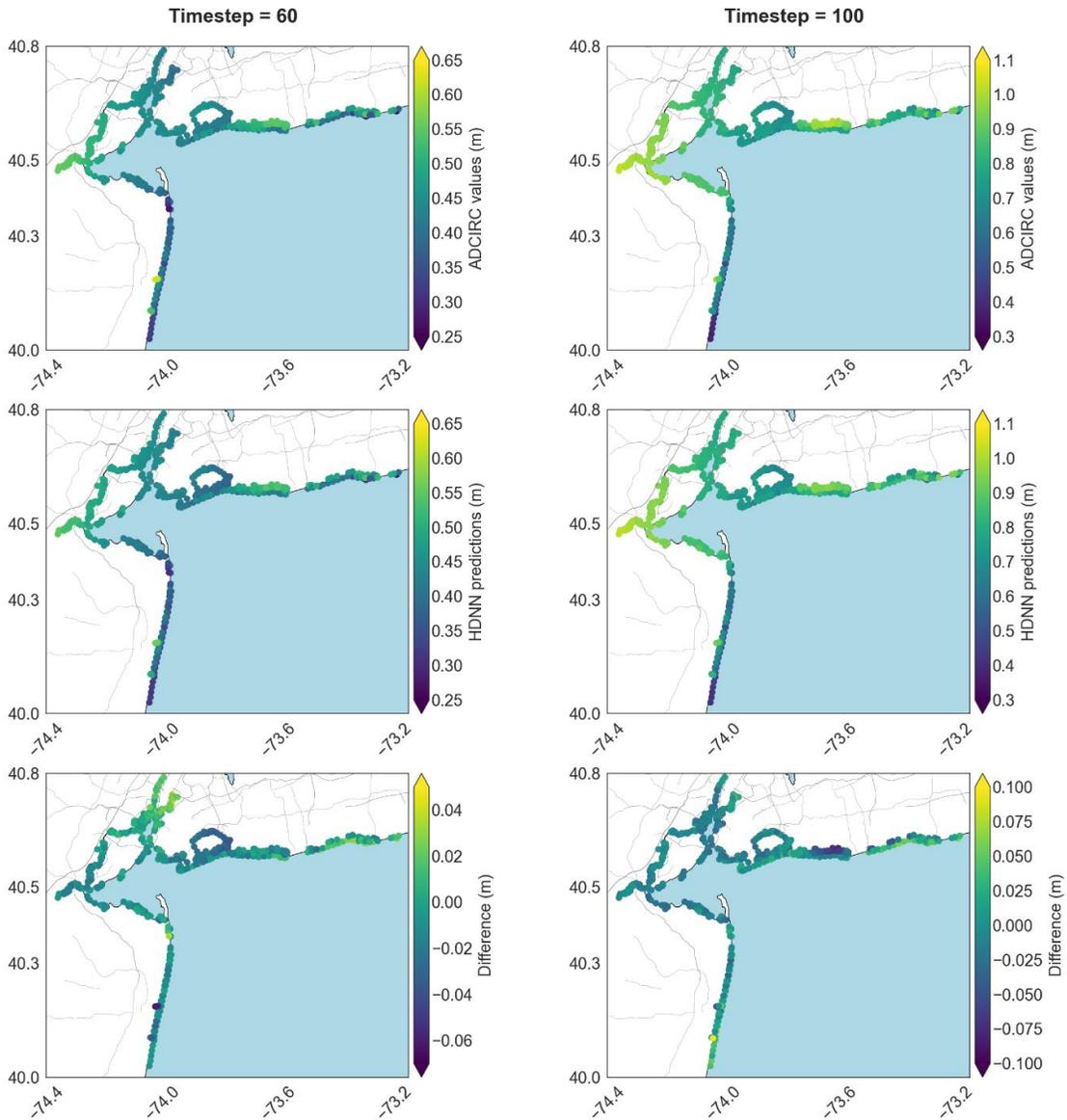

**Fig. 10.** High-dimensional storm surge predictions for storm ID 293 at time steps 60 and 100

## 4. Concluding remarks

In this study, the hierarchical deep neural networks (HDNNs) were developed along with convolutional autoencoders to accurately and efficiently simulate time-dependent storm surge across a large coastal area. This machine learning-based hybrid scheme leverages convolutional autoencoders to reduce the dimensionality of storm surge data by creating a latent space that captures key information, and then trains HDNNs on distinct time steps to map storm parameters (central pressure deficit, radius of maximum winds, storm latitude, and storm longitude) to this latent space. The HDNNs sequentially predicts future states, passing them as inputs to subsequent neural networks. The simulated low-dimensional representations are finally decoded back into storm surge time series. The proposed scheme was trained and evaluated using synthetic data from the North Atlantic Comprehensive Coastal Study (NACCS), covering a vast region with 576 coastal locations and 344 storms. Simulation results in terms of mean squared error (MSE), root mean squared error (RMSE), and mean absolute error (MAE) demonstrated high accuracy for both the convolutional autoencoder and HDNNs across training, validation, and testing sets. Due to its accuracy and computational efficiency, the proposed scheme is a promising tool for practical applications such as early warning systems and risk assessments for coastal flooding. Its robust predictive capabilities can enhance preparedness and mitigation strategies, contributing to the resilience of coastal regions.

**Acknowledgements**

This work was supported by the Natural Sciences and Engineering Research Council of Canada (NSERC) [grant number CRSNG RGPIN 2022-03492].

# References

Adeli, E., Sun, L., Wang, J., & Taflanidis, A. A. (2023). An advanced spatio-temporal convolutional recurrent neural network for storm surge predictions. *Neural Computing and Applications*, *35*(26), 18971–18987. https://doi.org/10.1007/s00521-023-08719-2

Al Kajbaf, A., & Bensi, M. (2020). Application of surrogate models in estimation of storm surge:A comparative assessment. *Applied Soft Computing*, *91*, 106184. https://doi.org/10.1016/j.asoc.2020.106184

Ayyad, M., Hajj, M. R., & Marsooli, R. (2022). Artificial intelligence for hurricane storm surge hazard assessment. *Ocean Engineering*, *245*, 110435. https://doi.org/10.1016/j.oceaneng.2021.110435

Bajo, M., & Umgiesser, G. (2010). Storm surge forecast through a combination of dynamic and neural network models. *Ocean Modelling*, *33*(1), 1–9. https://doi.org/10.1016/j.ocemod.2009.12.007

Bij de Vaate, I., Slobbe, D. C., & Verlaan, M. (2024). Mapping the spatiotemporal variability in global storm surge water levels using satellite radar altimetry. *Ocean Dynamics*, *74*(3), 169–182. https://doi.org/10.1007/s10236-023-01596-2

Calafat, F. M., Wahl, T., Tadesse, M. G., & Sparrow, S. N. (2022). Trends in Europe storm surge extremes match the rate of sea-level rise. *Nature*, *603*(7903), 841–845. https://doi.org/10.1038/s41586-022-04426-5

Chen, W.-B., Liu, W.-C., & Hsu, M.-H. (2012). Predicting typhoon-induced storm surge tide with a two-dimensional hydrodynamic model and artificial neural network model. *Natural Hazards and Earth System Sciences*, *12*(12), 3799–3809. https://doi.org/10.5194/nhess-12-3799-2012

Coch, N. K. (2014). Unique Vulnerability of the New York–New Jersey Metropolitan Area to Hurricane Destruction. *Journal of Coastal Research*, *31*(1), 196–212. https://doi.org/10.2112/JCOASTRES-D-13-00183.1

Giaremis, S., Nader, N., Dawson, C., Kaiser, C., Nikidis, E., & Kaiser, H. (2024). Storm surge modeling in the AI era: Using LSTM-based machine learning for enhancing forecasting accuracy. *Coastal Engineering*, *191*, 104532. https://doi.org/10.1016/j.coastaleng.2024.104532

Gori, A., Lin, N., Xi, D., & Emanuel, K. (2022). Tropical cyclone climatology change greatly exacerbates US extreme rainfall–surge hazard. *Nature Climate Change*, *12*(2), 171–178. https://doi.org/10.1038/s41558-021-01272-7

Hashemi, M. R., Spaulding, M. L., Shaw, A., Farhadi, H., & Lewis, M. (2016). An efficient artificial intelligence model for prediction of tropical storm surge. *Natural Hazards*, *82*(1), 471–491. https://doi.org/10.1007/s11069-016-2193-4

Igarashi, Y., & Tajima, Y. (2021). Application of recurrent neural network for prediction of the time-varying storm surge. *Coastal Engineering Journal*, *63*(1), 68–82. https://doi.org/10.1080/21664250.2020.1868736

Jelesnianski, C. P. (1972). *SPLASH:(Special Program to List Amplitudes of Surges from Hurricanes). I, Landfall storms*.

Jia, G., & Taflanidis, A. A. (2013). Kriging metamodeling for approximation of high-dimensional wave and surge responses in real-time storm/hurricane risk assessment. *Computer Methods in Applied Mechanics and Engineering*, *261–262*, 24–38. https://doi.org/10.1016/j.cma.2013.03.012

Jiang, W., Zhong, X., & Zhang, J. (2024). Surge-NF: Neural Fields inspired peak storm surge surrogate modeling with multi-task learning and positional encoding. *Coastal Engineering*, *193*, 104573. https://doi.org/10.1016/j.coastaleng.2024.104573

Jung, W., Taflanidis, A. A., Kyprioti, A. P., Adeli, E., Westerink, J. J., & Tolman, H. (2023). Efficient probabilistic storm surge estimation through adaptive importance sampling across storm advisories. *Coastal Engineering*, *183*, 104287. https://doi.org/10.1016/j.coastaleng.2023.104287

Kijewski-Correa, T., Taflanidis, A., Vardeman, C., Sweet, J., Zhang, J., Snaiki, R., Wu, T., Silver, Z., & Kennedy, A. (2020). Geospatial Environments for Hurricane Risk Assessment: Applications to Situational Awareness and Resilience Planning in New Jersey. *Frontiers in Built Environment*, *6*. https://doi.org/10.3389/fbuil.2020.549106

Kim, S.-W., Melby, J. A., Nadal-Caraballo, N. C., & Ratcliff, J. (2015). A time-dependent surrogate model for storm surge prediction based on an artificial neural network using high-fidelity synthetic hurricane modeling. *Natural Hazards*, *76*(1), 565–585. https://doi.org/10.1007/s11069-014-1508-6

Kyprioti, A. P., Irwin, C., Taflanidis, A. A., Nadal-Caraballo, N. C., Yawn, M. C., & Aucoin, L. A. (2023). Spatio-temporal storm surge emulation using Gaussian Process techniques. *Coastal Engineering*, *180*, 104231. https://doi.org/10.1016/j.coastaleng.2022.104231

Kyprioti, A. P., Taflanidis, A. A., Nadal-Caraballo, N. C., & Campbell, M. (2021). Storm hazard analysis over extended geospatial grids utilizing surrogate models. *Coastal Engineering*, *168*, 103855. https://doi.org/10.1016/j.coastaleng.2021.103855

Kyprioti, A. P., Taflanidis, A. A., Plumlee, M., Asher, T. G., Spiller, E., Luettich, R. A., Blanton, B., Kijewski-Correa, T. L., Kennedy, A., & Schmied, L. (2021). Improvements in storm surge surrogate modeling for synthetic storm parameterization, node condition classification and implementation to small size databases. *Natural Hazards*, *109*(2), 1349–1386. https://doi.org/10.1007/s11069-021-04881-9

Lee, J.-W., Irish, J. L., Bensi, M. T., & Marcy, D. C. (2021). Rapid prediction of peak storm surge from tropical cyclone track time series using machine learning. *Coastal Engineering*, *170*, 104024. https://doi.org/10.1016/j.coastaleng.2021.104024

Liou, C.-Y., Cheng, W.-C., Liou, J.-W., & Liou, D.-R. (2014). Autoencoder for words. *Neurocomputing*, *139*, 84–96. https://doi.org/10.1016/j.neucom.2013.09.055

Liu, Y., Kutz, J. N., & Brunton, S. L. (2022). Hierarchical deep learning of multiscale differential equation time-steppers. *Philosophical Transactions of the Royal Society A: Mathematical, Physical and Engineering Sciences*, *380*(2229), 20210200. https://doi.org/10.1098/rsta.2021.0200

Lockwood, J. W., Lin, N., Oppenheimer, M., & Lai, C.-Y. (2022). Using Neural Networks to Predict Hurricane Storm Surge and to Assess the Sensitivity of Surge to Storm Characteristics. *Journal of Geophysical Research: Atmospheres*, *127*(24), e2022JD037617. https://doi.org/10.1029/2022JD037617

Pachev, B., Arora, P., del-Castillo-Negrete, C., Valseth, E., & Dawson, C. (2023). A framework for flexible peak storm surge prediction. *Coastal Engineering*, *186*, 104406. https://doi.org/10.1016/j.coastaleng.2023.104406

Plumlee, M., Asher, T. G., Chang, W., & Bilskie, M. V. (2021). High-fidelity hurricane surge forecasting using emulation and sequential experiments. *The Annals of Applied Statistics*, *15*(1), 460–480. https://doi.org/10.1214/20-AOAS1398

Qin, Y., Wei, Z., Chu, D., Zhang, J., Du, Y., & Che, Z. (2024). Artificial neural network-based multi-input multi-output model for short-term storm surge prediction on the southeast coast of China. *Ocean Engineering*, *300*, 116915. https://doi.org/10.1016/j.oceaneng.2024.116915

Rajasekaran, S., Gayathri, S., & Lee, T.-L. (2008). Support vector regression methodology for storm surge predictions. *Ocean Engineering*, *35*(16), 1578–1587. https://doi.org/10.1016/j.oceaneng.2008.08.004

Ramos-Valle, A. N., Curchitser, E. N., Bruyère, C. L., & McOwen, S. (2021). Implementation of an Artificial Neural Network for Storm Surge Forecasting. *Journal of Geophysical Research: Atmospheres*, *126*(13), e2020JD033266. https://doi.org/10.1029/2020JD033266

Rao, N. S. B., & Mazumdar, S. (1966). A Technique for Forecasting Storm Waves. *MAUSAM*, *17*(3), Article 3. https://doi.org/10.54302/mausam.v17i3.5723

Saviz Naeini, S., & Snaiki, R. (2023). Machine Learning Approximation for Rapid Prediction of High-Dimensional Storm Surge and Wave Responses. In R. Gupta, M. Sun, S. Brzev, M. S. Alam, K. T. W. Ng, J. Li, A. El Damatty, & C. Lim (Eds.), *Proceedings of the Canadian Society of Civil Engineering Annual Conference 2022* (pp. 701–710). Springer International Publishing. https://doi.org/10.1007/978-3-031-34593-7_43

Saviz Naeini, S., & Snaiki, R. (2024a). A physics-informed machine learning model for time-dependent wave runup prediction. *Ocean Engineering*, *295*, 116986. https://doi.org/10.1016/j.oceaneng.2024.116986

Saviz Naeini, S., & Snaiki, R. (2024b). A novel hybrid machine learning model for rapid assessment of wave and storm surge responses over an extended coastal region. *Coastal Engineering*, *190*, 104503. https://doi.org/10.1016/j.coastaleng.2024.104503

Silvester, R. (1970). Computation of storm surge. In *Coastal Engineering 1970* (pp. 1995–2010).

Snaiki, R., & Wu, T. (2020). Hurricane Hazard Assessment Along the United States Northeastern Coast: Surface Wind and Rain Fields Under Changing Climate. *Frontiers in Built Environment*, *6*. https://doi.org/10.3389/fbuil.2020.573054

Sztobryn, M. (2003). Forecast of storm surge by means of artificial neural network. *Journal of Sea Research*, *49*(4), 317–322. https://doi.org/10.1016/S1385-1101(03)00024-8

Thomas, T. J., & Dwarakish, G. S. (2015). Numerical Wave Modelling – A Review. *Aquatic Procedia*, *4*, 443–448. https://doi.org/10.1016/j.aqpro.2015.02.059

Wetzel, S. J. (2017). Unsupervised learning of phase transitions: From principal component analysis to variational autoencoders. *Physical Review E*, *96*(2), 022140. https://doi.org/10.1103/PhysRevE.96.022140

Xie, W., Xu, G., Zhang, H., & Dong, C. (2023). Developing a deep learning-based storm surge forecasting model. *Ocean Modelling*, *182*, 102179. https://doi.org/10.1016/j.ocemod.2023.102179

Yu, X., & Wu, T. (2023). Simulation of unsteady flow around bluff bodies using knowledge-enhanced convolutional neural network. *Journal of Wind Engineering and Industrial Aerodynamics*, *236*, 105405. https://doi.org/10.1016/j.jweia.2023.105405

Zhang, J., Taflanidis, A. A., Nadal-Caraballo, N. C., Melby, J. A., & Diop, F. (2018). Advances in surrogate modeling for storm surge prediction: Storm selection and addressing characteristics related to climate change. *Natural Hazards*, *94*(3), 1225–1253. https://doi.org/10.1007/s11069-018-3470-1